\documentclass[a4paper,11pt]{article}
\pdfoutput=1 

\usepackage{jinstpub} 

\usepackage{multirow}
\usepackage{subcaption}

\newcommand{\mum}{~\rm \upmu \rm m}

\title{\boldmath Radiation damage caused by cold neutrons in boron doped CMOS active pixel sensors}

\author[a,1]{B. Linnik,}
\author[a]{T. Bus,}
\author[a]{M. Deveaux,\note{Corresponding author.}}
\author[a]{D. Doering,}
\author[a]{P. Kudejova,}
\author[b]{F. M. Wagner,}
\author[a]{A. Yazgili,}
\author[a]{J. Stroth}

\affiliation[a]{Goethe University Frankfurt,\\Max-von-Laue-Str. 1, 60438 Frankfurt/M, Germany}
\affiliation[b]{Forschungsneutronenquelle Heinz-Maier-Leibnitz (FRM II), Technical University Munich, Lichtenbergstr. 1, 85747 Garching, Germany }

\emailAdd{deveaux@physik.uni-frankfurt.de}

\abstract{CMOS Monolithic Active Pixel Sensors (MAPS) are considered as an emerging technology in the field of charged particle tracking. They will be used in the vertex detectors of experiments like STAR, CBM and ALICE and are considered for the ILC and the tracker of ATLAS. In those applications, the sensors are exposed to sizeable radiation doses. \\
While the tolerance of MAPS to ionizing radiation and fast hadrons is well known, the damage caused by low energy neutrons was not studied so far.
Those slow neutrons may initiate nuclear fission of $^{10}$B dopants found in the B-doped silicon active medium of MAPS. This effect was expected
to create an unknown amount of radiation damage beyond the predictions of the NIEL (Non Ionizing Energy Loss) model for pure silicon.\\
We estimate the impact of this effect by calculating the additional NIEL created by this fission. Moreover, we show first measured data for CMOS sensors which were irradiated with cold neutrons. The empirical results contradict the prediction of the updated NIEL model both, qualitatively and quantitatively: The sensors irradiated with slow neutrons show an unexpected and strong acceptor removal, which is not observed in sensors irradiated with MeV neutrons.}


\proceeding{8$^{\text{th}}$ International Workshop on Semiconductor Pixel Detectors for Particles and Imaging\\
  Monday, 5 September 2016 to Friday, 9 September 2016\\
  Sestri Levante}

\begin{document}
\maketitle
\flushbottom

\section{Introduction}
\label{sec:intro}

The ultra-light and highly granular CMOS Monolithic Active Pixel Sensors (MAPS) are considered as sensor technology for various charged particle detectors like the micro vertex detector (MVD) of the CBM experiment at FAIR\cite{PaperVertex2013}. A dedicated R\&D program of the PICSEL group of IPHC Strasbourg and the IKF Frankfurt aims to match their radiation tolerance to the requirements of the CBM-MVD, which amount $3 \cdot 10^{13}\rm~N_{eq}/cm^2$ and 3 Mrad per year of operation \cite{PaperVertex2015}.

\begin{figure}
  \begin{center}
   \begin{minipage}[c]{0.6\columnwidth}
   \includegraphics[width=0.95\columnwidth]{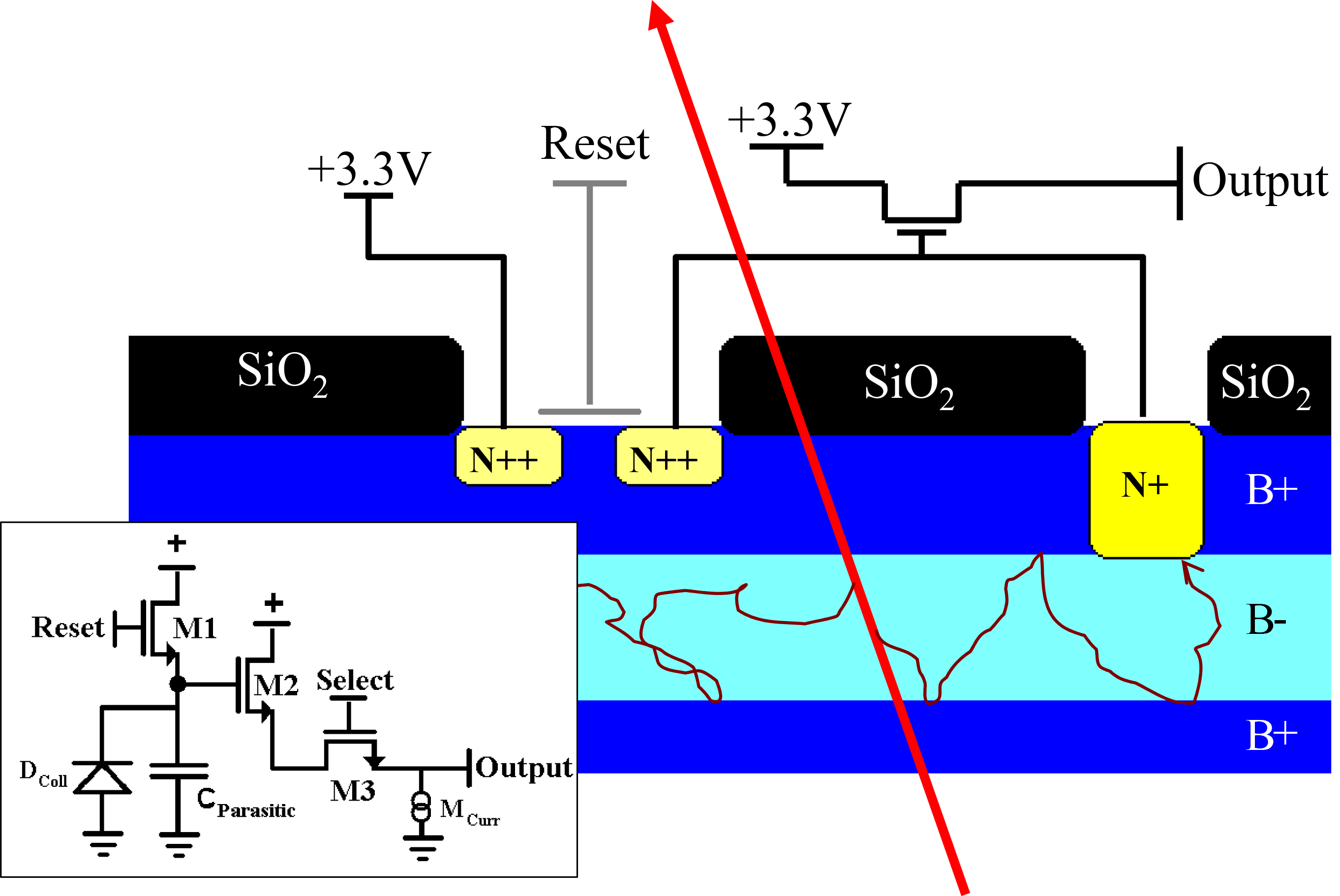} 
  \end{minipage} 
  \begin{minipage}[c]{0.35\columnwidth}
    \caption{Fundamental layout of a CMOS-sensor. The trajectory of an impinging minimum ionizing particle (red arrow) and the diffusion paths of two free electrons is shown. The schematics of the preamplifier of a 3T-pixel is shown in the lower left box.}
    \label{fig:SchematicsMAPS}
  \end{minipage}  
  \end{center}
\end{figure}

A 3T-pixel of a MAPS is shown in Figure \ref{fig:SchematicsMAPS}. It consists of three B-doped silicon layers: The highly doped substrate, a $\sim 15 \mum$ thick, moderately doped epitaxial layer and a thin, highly doped P-well with N-well implants. The partially depleted epitaxial layer serves as the active medium of the sensor. The N-well and the epitaxial layer form its diode. Free electrons, which were excited by particles impinging the epitaxial layer, diffuse in the epitaxial layer unless they reach the depleted volume of the diodes. The collected electrons discharge the parasitic capacity of the diode. The related voltage drop is buffered by a source follower and discriminated with an ADC/discriminator (not shown). The pixel capacity is recharged by opening the reset transistor.

The tolerance of MAPS to non-ionizing radiation damage has regularly been studied with fast \mbox{$\sim 1~\rm MeV$} reactor neutrons in accordance with the NIEL (Non-Ionizing-Energy-Loss) model \cite{NIEL-Modell}. This model assumes that the crystal damage caused by massive particles scales with their non-ionizing energy loss. The energy deposit is normalized to multiples of the one of $1~\rm MeV$ neutrons (N$_{\rm eq}$). Tables on the energy dependent NIEL of neutrons in pure silicon as plotted in Figure \ref{fig:hardnessfactorwithdoping} were used to estimate the required radiation tolerance of the sensors of the CBM-MVD. 

Additional NIEL is generated in B-doped silicon by a neutron induced fission of the boron dopants \mbox{(n + ${}^{10}$B $\rightarrow ^7$Li + $^4$He + 2.8~MeV)}. First radiation studies carried out with n-p-n transistors realized in the DMILL process \cite{BulkDamageInDMILL, RadiationDamageBipolar} suggest that this additional damage is sizable and may exceed the one expected from 1 MeV neutrons while it is considered as negligible in the standard NIEL model. 
To exclude a possible impact of thermal neutrons to the safe operation of the CBM-MVD, we studied the radiation damage caused by thermal neutrons in MAPS. This was done by estimating the additional radiation damage caused by the fission within the framework of the NIEL model. Moreover, we irradiated a MAPS prototype with cold neutrons and tested it for radiation damage.

%
%

\section{NIEL caused by neutron induced Boron fission, a theoretical estimate}
\begin{figure}[t]
  \centering
	\begin{subfigure}{.47\textwidth}
  \centering
	\includegraphics[width=1.0\linewidth]{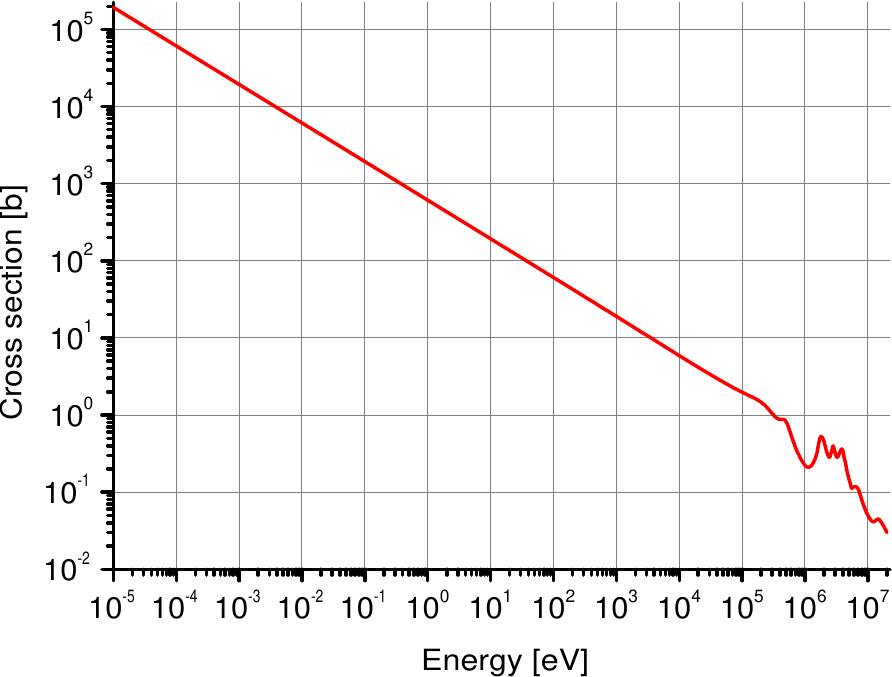}
  \caption{}
  \label{fig:B10Xsection}
\end{subfigure}%
\hspace*{0.02\textwidth}
\begin{subfigure}{.53\textwidth}
  \centering
  \includegraphics[width=1.0\linewidth]{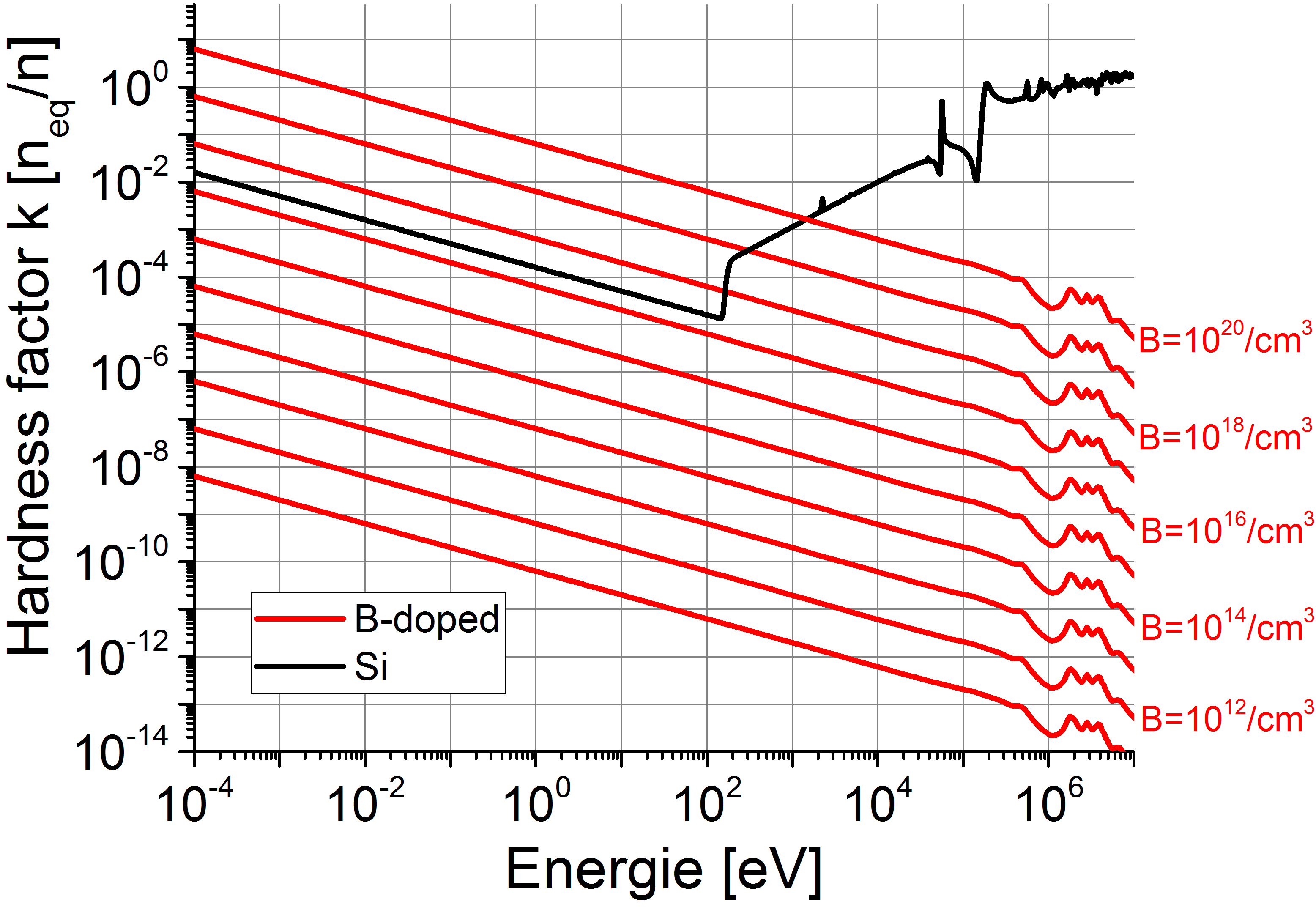}
  \caption{
  \label{fig:hardnessfactorwithdoping}}
\end{subfigure}
\caption{\newline (a) Cross section for the n+$^{10}$B$\rightarrow$ ${}^7$Li+${}^4$He +X reaction. Data from \cite{BorDataBase}. \\(b) Estimated hardness factor of neutrons impinging pure silicon (black, data from \cite{NIEL-compilation}), predicted damage due to boron fission (red). The prediction is not supported by data, see text.}
\end{figure}

Natural boron contains 19.8\% ${}^{10}$B and 80.2\% ${}^{11}$B. 
The neutron capture cross section of ${}^{11}$B is negligible, while the one of ${}^{10}$B amounts $\sim 3840~\rm b$ for thermal neutrons ($E= 0.025$ eV, see figure~\ref{fig:B10Xsection}). In the following, we assume that manufacturers of standard CMOS electronics use natural boron. The number of boron fissions per volume, $n_{\rm reactions}$, for monoenergetic neutrons with an energy $E$ is given by:
\begin{equation}
n_{\text{reactions}} =\sigma(E) \cdot I \cdot N_{\text{P}}\cdot \Phi_{\text{neutron}} 
\label{eqn:numberofreactionspervolume}
\end{equation}
Here, $\Phi_{\text{neutron}}$ is the integrated neutron flux in units of $\rm 1/cm^2$, $N_{\text{P}}$ the doping concentration. I=19.8\% is the isotopic factor of ${}^{10}$B and $\sigma(E)$ represents the energy dependent cross-section for the \mbox{n + ${}^{10}$B $\rightarrow ^7$Li + $^4$He + 2.8~MeV} reaction.
The energy of the impinging neutron and the frequent formation of a $\gamma$-ray with 478 keV is neglected. The 2.8 MeV fission energy provides $E_{\text{Li}}=1.1\text{~MeV}$ and $E_{\text{He}}=1.7\text{~MeV}$ to the fission products. The range of the ions was simulated with SRIM (Stopping Power of Ions in Matter) \cite{SRIM} and amounts in silicon to $6\mum$ and $3\mum$ for He and Li, respectively. Most energy is deposited in a Bragg peak at the end of the ion trajectory.

The ions deposit most of their energy by ionizing interactions. To estimate the non-ionizing energy deposit, we simulated the number of crystal vacancies caused by the ions with SRIM and compared them with the related number for 10~MeV protons, which have a known hardness factor. By doing so, we assume that the indicated number of vacancies scales with the hardness factor of the particles. This was tested for several proton energies between 100 keV and 10 MeV and a good agreement with a standard deviation of 7\% was observed. According to the simulation, the helium and lithium ions create in average 280 and 510 vacancies per ion, respectively. The sum amounts $V_B=790$ vacancies per boron fission. For 10~MeV protons (hardness factor $k_p=3.9$ N$_{\rm eq}$ \cite{summers}) penetrating a \mbox{10 $\mum$} silicon target, $V_P=684$ vacancies per cm trajectory were indicated. 


According to the NIEL model, the number of vacancies scales with the non-ionizing energy deposit. Provided the SRIM simulation is valid, one may compare the number of vacancies to estimate the NIEL of the fission products $D_B$ in units of $1~\rm MeV~ N_{eq}/cm^2$. One obtains:

\begin{equation}
D_B= \frac{V_B \cdot k_P}{V_P} \cdot \sigma(E) \cdot I \cdot N_P \cdot \Phi_{\rm neutron}=4.5 {~\rm cm} \cdot \sigma \cdot I \cdot N_P \cdot \Phi_{\rm neutron}
\end{equation}
The fission induced hardness factor for neutrons with a given energy is:
\begin{equation}
k_{\rm B} = 4.5 {~\rm cm}  \cdot \sigma \cdot I \cdot N_P
\end{equation}
Figure \ref{fig:hardnessfactorwithdoping} which compares $k_B$ of different doping concentrations with 
the hardness factor for pure silicon. One observes that the NIEL caused by the boron fission reaches the NIEL expected for pure silicon starting from doping concentrations slightly above $\sim 10^{17}\rm/cm^3$. 

The epitaxial layers of typical CMOS sensors have a doping concentration below $\lesssim 10^{15}\rm /cm^3$, the wells may reach $\sim 10^{17}\rm/cm^3$,
and the substrate shows typically a doping of $\sim 10^{19}\rm/cm^3$. Therefore, in first order, boron is not expected to cause additional radiation damage
in the MAPS. However, fission ions created in the highly doped structures may travel toward the epitaxial layer and cause bulk damage there. According to previous studies \cite{MichaelPHD}, this bulk damage may increase the leakage currents of the diodes and moreover reduce the charge collection efficiency of the sensor.

\section{Effects of cold neutrons on CMOS sensors}
To investigate the effects of boron fission, we obtained MIMOSA-19 sensors from the PICSEL group of IPHC Strasbourg and irradiated them at the FRM II research reactor in Garching. The sensors were manufactured in a 0{.}35$\mum$ CMOS technology and feature a 14$\mum$ thick epitaxial layer with an anticipated B-doping of few $10^{15}/\rm cm^3$. The $196\times196$ 3T-pixels with 12x12$\mum^2$ pitch have L-shaped diodes with 39.6$\mum^2$ surface. The anticipated doping of the substrate of the P-Well is $\sim10^{19}\rm/cm^3$. 

The chips were irradiated at the MEDAPP beam line~\cite{MEDAPP} with fast neutrons (a direct fission spectrum with $E_{\rm neutron}=2~\rm MeV$ (peak), hardness factor for pure Si $k\approx 1$ N$_{\rm eq}$, $E_{\rm neutron}>100~\rm keV$ for 99\% of all neutrons). Unwanted $\gamma$-rays caused an ionizing dose of \mbox{$<100~\rm krad$} per $10^{13} ~\rm n/cm^2$. Other sensors were irradiated at the PGAA beam line \cite{PaperPgaa} with cold neutrons (average energy $E_{\rm neutron} = 1.8\times 10^{-3}~\rm eV$ corresponding\footnote{As obtained from folding the spectrum given in \cite{PaperPgaa2} with hardness factors according to \cite{NIEL-compilation}.} to $k$=0.003 N$_{\rm eq}$) and an unknown ionizing dose. The irradiation was done at room temperature. Despite the sensors remained unpowered during irradiation, the ionizing radiation damage is considered to determine the leakage currents after irradiation.

\begin{figure}[t]
\centering
\begin{subfigure}{.5\textwidth}
  \centering
  \includegraphics[width=1.0\linewidth]{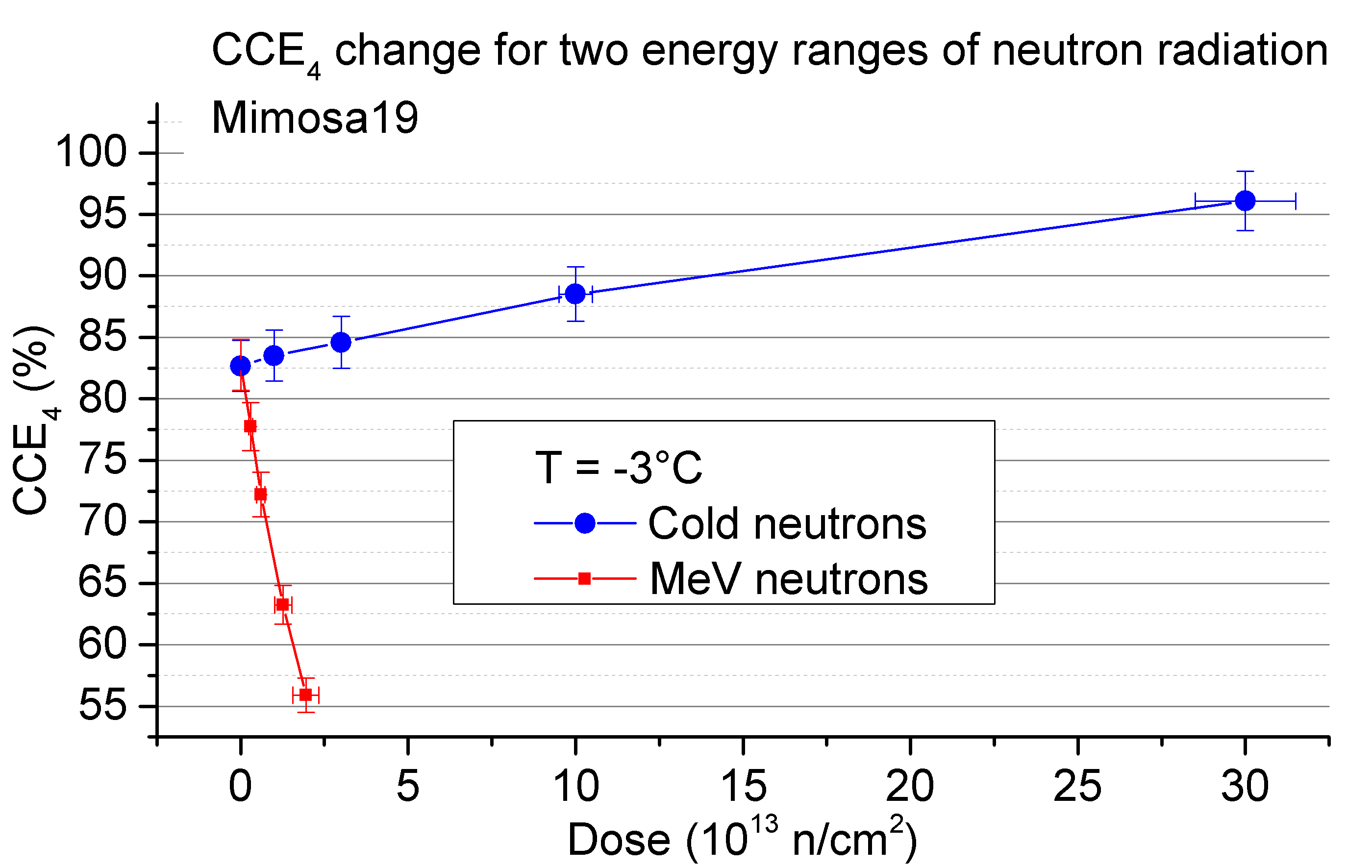}
  \caption{The charge collection efficiency for the four most significant pixel in a 5x5 cluster.}
  \label{fig:CCEthermal}
\end{subfigure}%
\hspace*{0.05\textwidth}
\begin{subfigure}{.5\textwidth}
  \centering
  \vspace{-0.5cm}
	\includegraphics[width=1.0\linewidth]{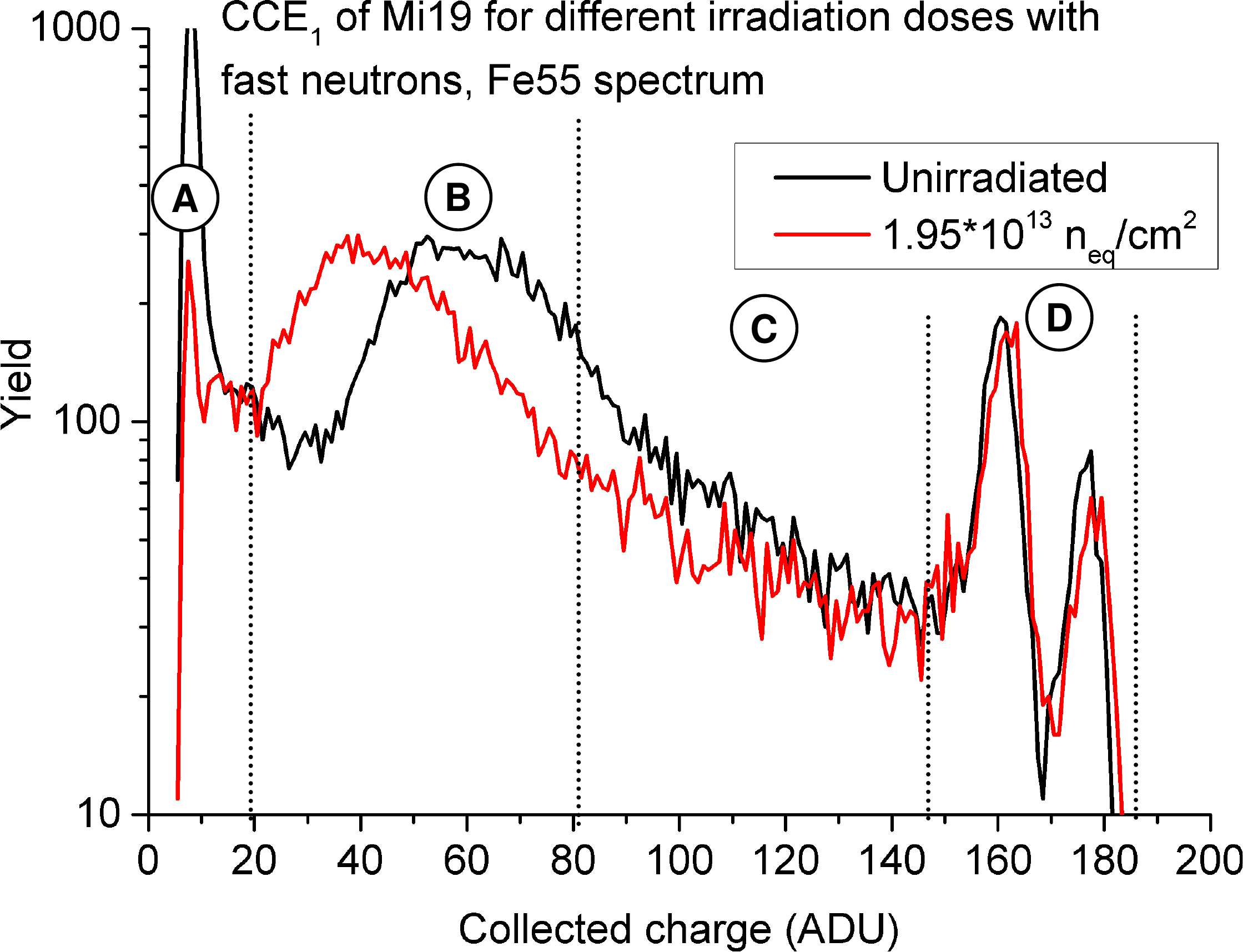}
  \caption{Amplitude spectrum of a reference sensor and a sensor irradiated with fast neutrons.
  \label{fig:CCE1Fast}}
\end{subfigure}
\newline
\begin{subfigure}{.5\textwidth}
  \centering
	\vspace{-0.8cm}
  \includegraphics[width=1.0\linewidth]{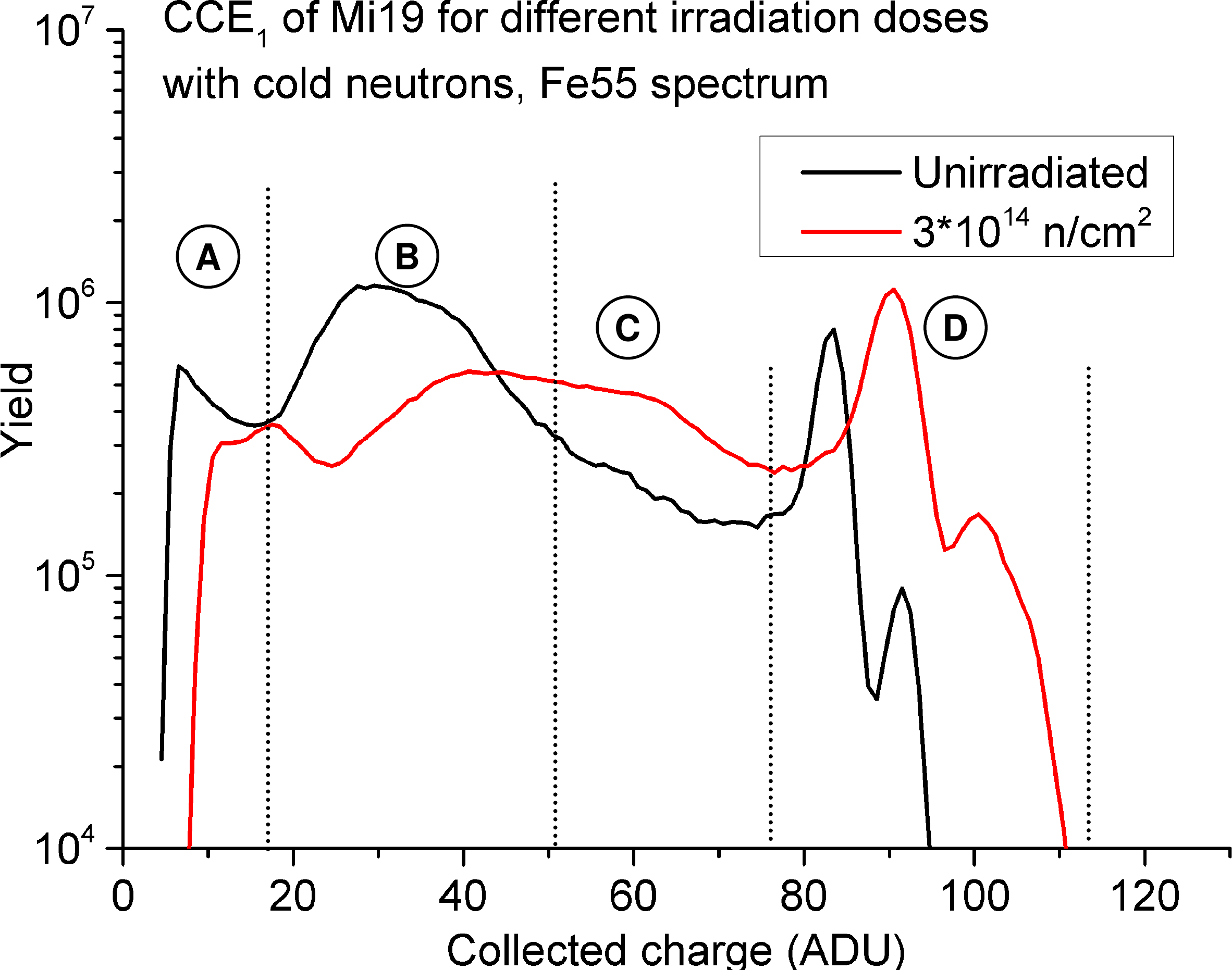}
  \caption{Amplitude spectrum of a reference sensor and a sensor irradiated with cold neutrons.
  \label{fig:CCE1Thermal}}
\end{subfigure}
\hspace*{0.02\textwidth}
\begin{subfigure}{.469\textwidth}
  \centering
    \vspace{0.8cm}
	\includegraphics[width=1.0\linewidth]{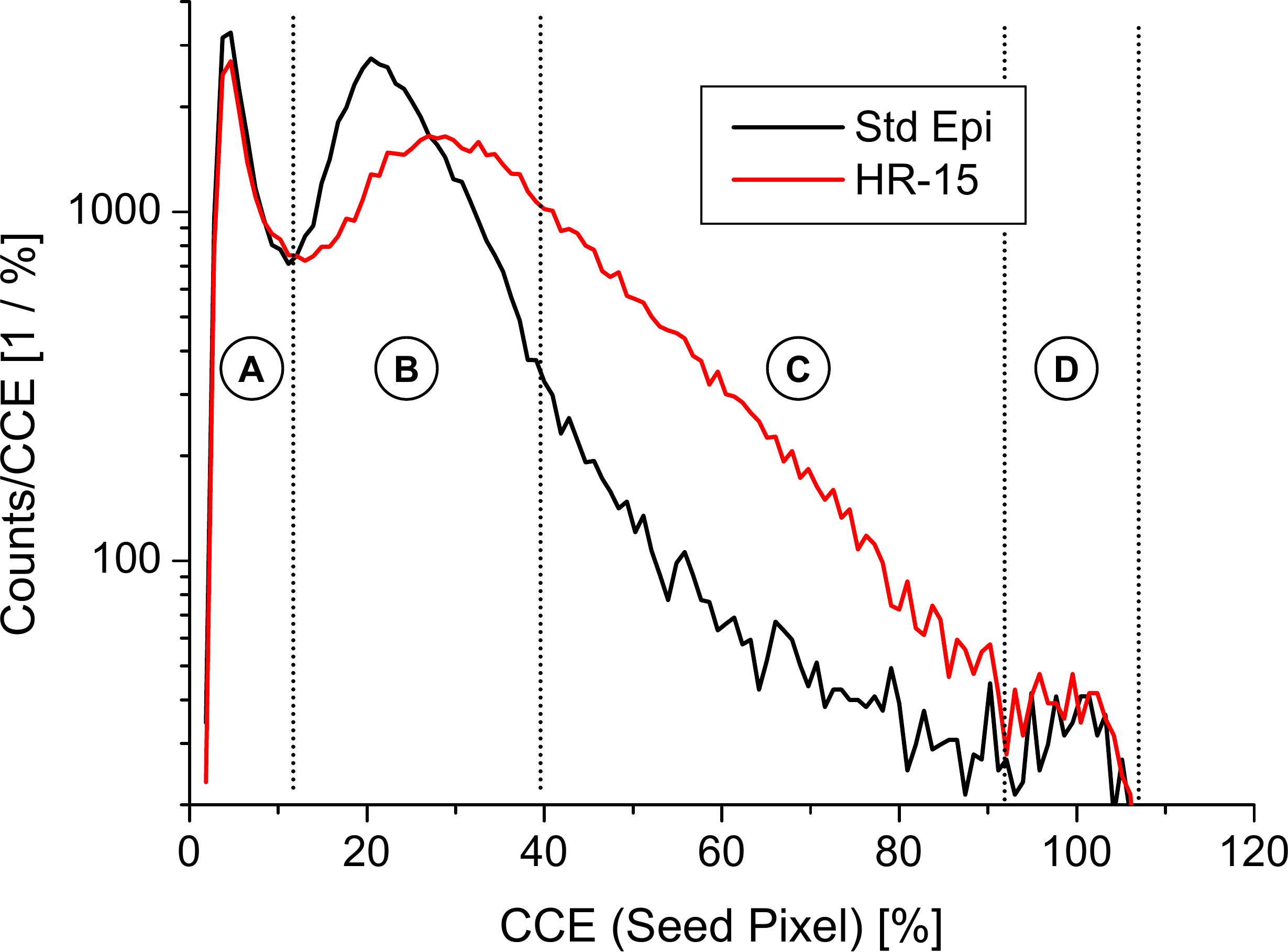}
  \caption{Comparison of the amplitude spectrum of a MIMOSA-26 sensor with highly (Std Epi) and lowly (HR-15) doped epitaxial layer. From \cite{RadToleranceDev}.  \label{fig:CCEMi26}}
\end{subfigure}
\caption{Charge collection of neutron irradiated MAPS. The amplitude spectra for intermediate doses follow the trend and were removed from panel \ref{fig:CCE1Fast} and \ref{fig:CCE1Thermal} for the sake of clarity.}
\end{figure}

The sensors were bonded, illuminated with X-ray sources ($^{55}$Fe$\rightarrow 5.9~\rm keV$, $^{109}$Cd$\rightarrow \rm 22.1~keV$) and their amplitude spectra were recorded according to our standard protocol (see \cite{MichaelPHD}). The summed amplitude spectrum of groups four pixels of X-ray clusters are known to provide a good relative measure of the charge collection efficiency (CCE) of neutron irradiated sensors. The CCE of MAPS with 3T-pixel was previously found not to be significantly modified by ionizing radiation damage (see e.g. \cite{MichaelPHD}, section 5.3.3 and 5.3.4.3). Comparing CCE measurements done with $^{55}$Fe and $^{109}$Cd was confirmed to be appropriate due to the linear response of the sensor.

The results of the study are displayed in Figure \ref{fig:CCEthermal}. The sensors irradiated at MEDAPP display a \emph{decrease} of the average CCE with increasing radiation dose, which matches the results of other studies (e.g. \cite{RadToleranceDev}). In the sensors irradiated at PGAA, one observes an unexpected \emph{increase} of the CCE. The slope of the CCE does thus depend on the neutron energy, what cannot be explained within the NIEL model. 

In order to obtain a closer understanding of the effect, we analyzed the amplitude spectra of the sensors in detail. For this purpose, we employed spectra, which show the charge deposited in the seed pixel of the hit clusters. Such a spectrum (see Figure \ref {fig:CCE1Fast}) may be subdivided into four regions: Region (b) reflects X-rays, which hit the non-depleted epitaxial layer. Region (c) shows the hits in an intermediate zone nearby the depleted volume of the diodes. Region (d) shows the response of the sensor to direct hits in the depleted volume. The low energy region (a) can be ignored here.

In sensors irradiated at MEDAPP with fast neutrons (Figure \ref {fig:CCE1Fast}), one observes region (d) to remain unchanged within uncertainties after irradiation. Therefore, the properties of the depleted volume is considered as unchanged. The peak in (b) is shifted to lower values after irradiation, which indicates a partial recombination of diffusion signal charge due to bulk damage.

For the sensor irradiated at PGAA with cold neutrons (Figure \ref {fig:CCE1Thermal}), region (b) is mostly depleted after irradiation and the remaining peak is shifted to higher values. Most of the entries missing in (b) are found in \mbox{region (c).} This indicates that the pixel diode improved its CCE for charge created in the nearby epitaxial layer and that there is no significant charge recombination after irradiation. The number of entries in region (d) increases significantly after irradiation. This is interpreted as an extension of the depleted volume of the pixel diode, which explains the previous observation. Moreover, the peak is shifted toward higher values. This suggests a reduced diode capacity as caused by an extended depleted volume. 

The findings are not explained by a radiation induced modification of the gain of the amplifiers of the readout chain. A linear modification would shift the peaks in (d) as observed. However, the number of entries in the peaks should remain constant, which does not match the observations. An radiation induced, non-linear modification of the amplifier response curve is also not suited to explain the modified shape of the spectrum. It this scenario, the peaks in region (d) would be smeared out\footnote{The dark signal of the pixels (before performing correlated double sampling) is widely spread. A non-linear amplifyer response would thus turn into a sizeable spread of gain from pixel to pixel. The formation of the peaks observed requires the pixels to show an identical gain.}, which is not observed.

A plausible explaination considers a strong, radiation induced acceptor removal, which reduces the doping of the epitaxial layer and such increases the size of the depleted zone of the diode. The numbers of entries in the peaks in region (d) increases by a factor slightly above two. For flat PN-junctions, this would turn into a drop in doping by a rough factor of five. 

The hypothesis of an acceptor removal is supporting the results of previous studies \cite{RadToleranceDev}, which compared the spectra of MIMOSA-26 sensors with known highly (Std Epi, few $10^{15}~\rm P/cm^3$) and lowly  doped (HR-15, few $10^{13}~\rm P/cm^3$) epitaxial layer \cite{Mi26Ref}. As shown in Figure \ref{fig:CCEMi26}, the reduced doping created similar modifications of the amplitude spectrum in region (b) and (c). A significant increase of the depleted volume -region (d)- was not observed with MIMOSA-26, which was interpreted as a consequence of the small size and depletion voltage of the diode of this chip.

Note that, unless isotope enriched boron was used for doping, the strong acceptor removal observed cannot be explained with the physical destruction of the ${}^{10}$B atoms alone. This is as natural boron contains 80.2\% of ${}^{11}$B, which shows only a small fission cross section and thus should remain mostly stable. We investigated if  transmutation N-doping (a phosphorus production by thermal neutrons in silicon, n + ${}^{30}$Si $\rightarrow ^{30}$P + e + $\overline{\nu}_e$) could explain our findings. The number of $^{30}$P atoms produced was estimated to about three orders of magnitude below the number of missing acceptors, which seems to exclude this scenario. It might be worth speculating, if the slow ions created in the ${}^{10}$B-fission may each produce numerous traps, which act as effective N-dopands.

\section{Summary and conclusion}
The response of B-doped silicon sensors to thermal radiation has been studied theoretically and experimentally. Computations based on the NIEL model predict that the neutron induced fission of boron dopands in silicon (n + ${}^{10} $B $\rightarrow ^7$Li + $^4$He + 2.8~MeV) will add significant bulk damage in case the doping concentration exceeds $p=10^{17}/\rm cm^3$. Ions being created in the P-wells and the substrate were expected to damage the active volume of MAPS, which is by itself too lowly doped to obtain significant additional damage. We expected therefore that cold neutrons would reduce the charge collection efficiency (CCE) of sensors exposed to cold neutrons to some extent. This prediction is proven wrong by our observations made on sensors irradiated with cold neutrons. We observe an excessive acceptor removal, which is not observed in sensors irradiated with fast neutrons, and increases the CCE of the sensor. This suggests that cold and fast neutrons generate traps with different properties. The traps created by the fast neutrons do dominantly reduce the charge carriers lifetime. The traps created by thermal neutrons create dominantly a substantial acceptor removal and the reduction of charge carriers lifetime is either absent or compensated by the faster charge collection found in lowly-doped epitaxial layers.

We conclude that the sensor reacts in a fundamentally different way to slow and fast neutrons. Therefore, the NIEL model appears not suited to parametrize or predict the radiation damage induced by slow neutrons to our sensor. As MIMOSA-19 does not allow for a direct access of the doping concentration, e.g. by C-V measurements of the diode, our findings remain somewhat qualitative. We recommend therefore to repeat the study with suited test structures.

 \acknowledgments
This work was supported by HIC for FAIR, GSI and the BMBF (06FY90991, 05P12RFFC7).

\end{document}